\documentclass[aps,prl,twocolumn,amsmath,amssymb,superscriptaddress,reprint]{revtex4-1}

\usepackage{amsmath,mathtools}
\usepackage{graphicx}
\usepackage[dvipsnames]{color}
\usepackage{hyperref}
\usepackage{psfrag}
\usepackage{stmaryrd}
\usepackage{amssymb}
\usepackage{wasysym}
\usepackage{float}
\usepackage{color}

\usepackage[latin1]{inputenc}
\usepackage[T1]{fontenc}
\usepackage[french,english]{babel}
\usepackage{lmodern}

\begin{document}

\title{Quantitative experimental observation of weak inertial-wave turbulence}

\author{Eduardo Monsalve}
\affiliation{Université Paris-Saclay, CNRS, FAST, 91405, Orsay, France}
\author{Maxime Brunet}
\affiliation{Université Paris-Saclay, CNRS, FAST, 91405, Orsay, France}
\author{Basile Gallet}
\affiliation{Université Paris-Saclay, CNRS, CEA, Service de Physique de  
l'\'Etat Condensé, 91191, Gif-sur-Yvette, France}
\author{Pierre-Philippe~Cortet}
\email[]{ppcortet@fast.u-psud.fr} \affiliation{Université Paris-Saclay, CNRS, FAST, 91405, Orsay, France}

\date{\today}

\begin{abstract}
We report the quantitative experimental observation of the weak inertial-wave 
turbulence regime of rotating turbulence. We produce a statistically steady 
homogeneous turbulent flow that consists of nonlinearly interacting inertial 
waves, using rough top and bottom boundaries to prevent the emergence of a 
geostrophic flow. As the forcing amplitude increases, the temporal spectrum 
evolves from 
a discrete set of peaks to a continuous spectrum. Maps of the bicoherence of 
the velocity field confirm such a gradual transition between discrete wave 
interactions at weak forcing amplitude, and the regime described by weak 
turbulence theory (WTT) for stronger forcing. In the former regime, the 
bicoherence maps display a near-zero background level, together with sharp 
localized peaks associated with discrete resonances. By contrast, in the latter 
regime the bicoherence is a smooth function that takes values of the order of 
the Rossby number, in line with the infinite-domain and random-phase 
assumptions of WTT. The spatial spectra then display a power-law behavior, both 
the spectral exponent and the spectral level being accurately predicted by WTT 
at high Reynolds number and low Rossby number.
\end{abstract}

\maketitle

\textit{Introduction.---} Wave turbulence theory (WTT) addresses the statistical
properties of weakly nonlinear ensembles of waves in large
domains~\cite{Zakharov1992,Newell2011,Nazarenko2011}. The theory provides a
rigorous analytical framework to derive quantitative predictions for the kinetic
energy spectrum, a task that remains extremely challenging for standard
hydrodynamic turbulence. WTT appears of utmost interest for 3D fluid systems in
which bulk waves can propagate, such as rotating or stratified
fluids~\cite{Pedlosky1987,Davidson2013} where such quantitative predictions
could pave the way for better turbulence parameterizations in coarse atmospheric
and oceanic models~\cite{Gregg}. WTT has already proven a valuable conceptual
tool to understand energy transfers in 2D wave systems, such as surface gravity
or capillary waves~\cite{Falcon2007,Clark2014,Aubourg2016,Berhanu2019} and
bending waves in thin elastic plates~\cite{Cobelli2009,Humbert2013,Miquel2013}. 
Comparatively, 3D fluid systems present additional
complications that have hindered progress at the experimental and numerical
level. For instance, rapidly rotating fluids support inertial waves associated 
with the restoring action of the Coriolis force~\cite{Greenspan1968}, but these 
waves represent only a subset of the possible fluid motions. The
emerging slow geostrophic flows (wandering vortices invariant along the 
rotation axis, denoted as the vertical axis in the following) are not included 
in 
WTT~\cite{Galtier2003,Nazarenko2011b,Galtier2020}, and yet they
represent a significant fraction of the kinetic energy in most rotating
turbulence experiments~\cite{Hopfinger1982,Yarom2013,Campagne2015} and numerical
simulations~\cite{Sen2014,Clark2014b,Deusebio2014}. Geostrophic turbulence then
interacts with wave turbulence, advecting and distorting the wave field
\cite{Clark2014b,Campagne2015}. Part of the reason for the emergence of strong
geostrophic flows is that most experiments were driven using standard 
forcing mechanisms of hydrodynamic turbulence --
grids~\cite{Jacquin1990,Staplehurst2008,Lamriben2011},
propellers~\cite{Campagne2016}, jets~\cite{Baroud2003,Yarom2013} -- which
project poorly onto the spatio-temporal structure of inertial waves. However,
even when the flow is driven in such a way as to induce inertial waves only, it
was recently shown that geostrophic flows arise spontaneously through
instability processes~\cite{LeReun2019,Brunet2020}.

Because of these difficulties, recent advances on inertial wave turbulence have
consisted in detecting waves coexisting with 2D geostrophic flows
\cite{Yarom2014,Clark2014b,Campagne2015,Yarom2017}, isolating regimes of wave 
dynamics in
the absence of 2D flows \cite{LeReun2017,Brunet2020}, and providing evidence of
the triadic resonance instability (TRI)~\cite{LeReun2017,LeReun2019,Brunet2020}.
Building on these previous works, we report on an experimental
study of a spatially homogeneous weakly nonlinear inertial-wave-driven flow in
which we can test the quantitative predictions of WTT. We focus on the
prediction for the spatial spectrum in statistically steady state, the central
object of WTT. According to Refs.~\cite{Galtier2003,Nazarenko2011b}, weak
inertial-wave turbulence consists of energy transfers towards low frequencies
and small horizontal scales, while the vertical scale remains comparable to the
vertical injection scale $L_\parallel$. The flow is described as a set of 
inertial waves that interact nonlinearly through triadic interaction 
coefficients (see e.g. Eq.~(4) from the Supplemental Material~\cite{SM}). In 
the asymptotic limit $k_\perp L_\parallel \gg 1$ the 
triadic interaction coefficients admit an $L_\parallel$-independent limit, so 
that $\Omega$ and $L_\parallel$ enter the equations only through oscillatory 
factors involving the wave frequencies. Because the wave frequencies are 
proportional to $\Omega/L_\parallel$, we conclude that only 
$\Omega/L_\parallel$ should enter dimensional analysis in that limit, instead 
of $\Omega$ and $L_\parallel$ independently. Furthermore, inertial-wave 
turbulence proceeds through three-wave interactions, for which WTT
predicts that the energy spectrum $E(k_\perp)$ is proportional to the square
root of the energy flux $\epsilon$~\cite{Connaughton,Nazarenko2011}. 
Dimensional
analysis using $E(k_\perp) \epsilon^{-1/2}$, $\Omega / L_\parallel$, and
$k_\perp$ then yields: \begin{equation} E(k_\perp) = {\cal C} \,
\sqrt{\frac{\epsilon \Omega}{L_\parallel}} k_\perp^{-5/2} \, ,
\label{WTTspectrum} \end{equation} where ${\cal C}$ is a dimensionless 
constant. This dimensional argument can be made rigorous within the WTT 
framework~\cite{Galtier2003}, which in principle provides the value of the 
constant ${\cal C}$.
WTT thus predicts the exponent of the velocity spectrum in the self-similar
regime associated with a forward energy cascade, together with the dependence of
the spectral level on the global rotation rate $\Omega$ and the cascading 
energy flux $\epsilon$.

In the following, we introduce the experimental setup, before assessing the
validity of the assumptions of WTT in the present experiments. We then confront
the prediction~(\ref{WTTspectrum}) to the experimental data. The two appear to 
be
in excellent agreement, both in terms of spectral exponent and spectral level.

\begin{figure} 
	\centerline{\includegraphics[width=7.5cm]{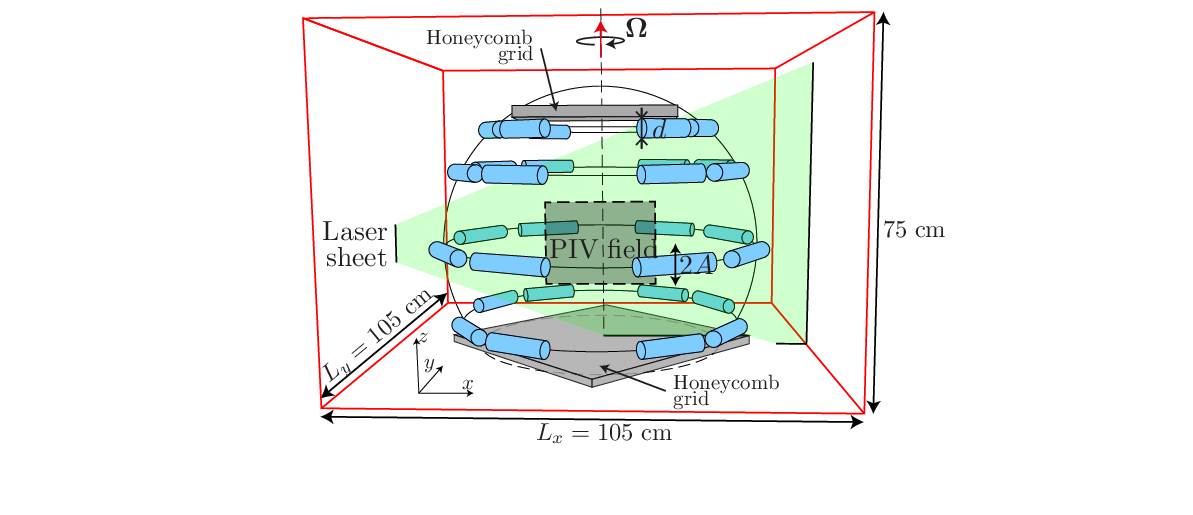}}
	\caption{Experimental setup. 32 horizontal cylinders are tangent to an
	80-cm-diameter virtual sphere horizontally centered in a water tank 
	mounted on a rotating platform. The flow is bounded by two honeycomb grids 
	at top and bottom (grey plates). The cylinders oscillate vertically at angular 
	frequency~$\omega_0$.\label{fig:setup}}
\end{figure}

\textit{Experimental setup.---} The experimental setup is sketched in
Fig.~\ref{fig:setup}. It is an evolution of the setup described in Brunet
\textit{et al.}~\cite{Brunet2020}, and we only recall its salient features. $32$
horizontal cylinders of diameter $d=4$~cm and length between $12$ and $18$~cm
oscillate vertically inside a parallelepipedic tank of $105\times 105$~cm$^2$
base filled with $63$~cm of water. The cylinders are arranged regularly around
an 80-cm-diameter virtual sphere horizontally centered in the water tank. The 
virtual sphere is truncated by the bottom of the tank, allowing us (i) to 
consider a sphere diameter that is greater than the water depth, and (ii) to 
take advantage of turbulent friction on the bottom boundary (as described 
below). Each
cylinder follows a vertical sinusoidal oscillatory motion of amplitude $A$ and 
angular frequency $\omega_0$, with independent random initial phases for the 
$32$ cylinders.

The entire apparatus is mounted on a 2-m-diameter platform rotating at a rate
$\Omega$ ranging from $4.5$ to $18$~rpm around the vertical axis $z$. The
cylinders oscillate at angular frequency $\omega_0 = 0.84 \times 2\Omega$,
generating self-similar inertial-wave beams~\cite{Cortet2010,Machicoane2015}
that spread as they propagate. The 
precise value $\omega_0/2\Omega=0.84$ is arbitrary and not crucial to our 
results. Nevertheless, it conveniently leads to most of the wave beams 
propagating toward the central region of the tank. The
amplitude $A$ of oscillation of the cylinders ranges from a few millimeters to
$25$~mm, leading to a forcing Reynolds number $270\leq Re=A\omega_0 d/\nu \leq
3080$ for $\Omega=18$~rpm. The forcing Rossby number
$A\omega_0/2\Omega d$ varies in the range $0.05\leq A\omega_0/2\Omega d \leq
0.52$ for $\Omega=18$~rpm (for $\Omega=4.5$~rpm, $250\leq Re \leq 710$ and
$0.17\leq A\omega_0/2\Omega d \leq 0.48$). As the forcing amplitude increases,
the overlapping wave beams generated by the 32 wavemakers produce a nearly
statistically homogeneous flow in the central region (see movies of the 
velocity field in the Supplemental Material~\cite{SM}).

A crucial modification to the previous version of the apparatus is the addition
of two horizontal honeycomb grids (2.5~cm in height, 2.7~cm in mesh), one at 
the bottom of the tank and one at
$59$~cm from the bottom (see Fig.~\ref{fig:setup}). As shown in 
Ref.~\cite{Brunet2020}, a single
honeycomb grid efficiently damps geostrophic motion through enhanced 
turbulent drag on the rough grid topography, with little impact on wave
dynamics. In the present study, we have included a second such grid to fully
suppress spontaneous energy transfers to geostrophic modes, in
a similar fashion to the numerical study of Le Reun \textit{et 
al.}~\cite{LeReun2017}. We have also upgraded the wave-driving mechanism to
increase the maximum $Re$ by a factor of three.

Two components $(u_x,u_z)$ of the velocity field are measured in a vertical
plane containing the center of the virtual sphere using a double-frame particle
image velocimetry (PIV) system mounted on the rotating platform. The velocity
fields have a spatial resolution of $1.93$~mm over an area of $\Delta x \times
\Delta z = 285 \times 214$~mm$^2$ at the center of the virtual sphere
(Fig.~\ref{fig:setup}). For each experimental run, PIV acquisition covers 
$1250$~periods of the forcing in the statistically steady flow regime.

\begin{figure}
	\centering
	\includegraphics[width=8cm]{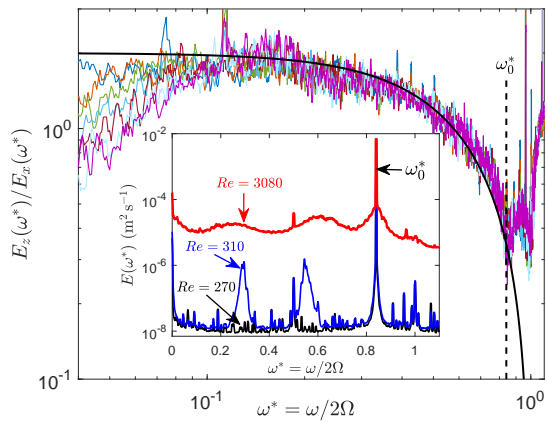} 
	\caption{Ratio of the power spectral densities of the vertical and 
		horizontal components of the velocity, for the
		experiments at $\Omega=18$~rpm and $Re = 560, 700, 1060, 1540, 2020, 2540$ and $3080$. The solid line 
		$E_z/E_x=2(1-{\omega^*}^2)/(1+{\omega^*}^2)$ shows 
		the behavior expected for an axisymmetric distribution of inertial 
		waves. \textbf{Inset:} Power spectral density $E(\omega^*)$ of the 
		measured velocity field, for $\Omega = 18$~rpm and three $Re$. 
		\label{fig:comp_anisot_18_nid}}
\end{figure}

\textit{Temporal dynamics.---} In the inset of
Fig.~\ref{fig:comp_anisot_18_nid}, we show the temporal power spectral density
$E(\omega^*=\omega/2\Omega)$ of the measured velocity field for $\Omega=18$~rpm 
and three values of $Re$. For the lowest forcing amplitude $Re=270$, the 
spectrum is dominated by a 
peak at normalized
frequency $\omega_0^*=\omega_0/2\Omega=0.84$ corresponding to the waves forced 
by the wavemakers. The spectrum at $Re=310$ displays two
additional subharmonic peaks, for two frequencies $\omega_1^*\simeq 0.29$ and
$\omega_2^*\simeq 0.55$ in triadic resonance with the forcing frequency:
$\omega_1^*+\omega_2^*=\omega_0^*$. These secondary peaks result from the TRI 
of 
the wave beams generated by the
forcing~\cite{Bordes2012}, i.e., the very first stage of nonlinear energy
transfers between the base flow and other frequencies and spatial scales (TRI 
criteria for wave beams are discussed in Refs.~\cite{Bourget2014,Fan2020}).
Finally, the spectrum at $Re=3080$ illustrates the regime of developed
turbulence, where the flow has populated a continuous range of frequencies.

In the context of rotating turbulence, a 
natural question to ask is whether this energy is carried by inertial waves or 
geostrophic eddies. Indeed, in Brunet~\textit{et al.}~\cite{Brunet2020}, we
showed that the waves could 
spontaneously transfer energy to low-frequency geostrophic vortex modes. The 
absence of an energetic peak at $\omega=0$ in the spectra 
already indicates that such energy condensation into a 
geostrophic mode is suppressed by the honeycomb grids. An efficient way to
test whether the spectral content corresponds to inertial 
waves consists in computing the ratio $E_z/E_x$ of the power spectral densities 
$E_z(\omega^*)$ and $E_x(\omega^*)$ of the horizontal and vertical velocity
components, respectively. Indeed, for a single
inertial wave, the ratio of the amplitudes of oscillation of the vertical and
horizontal velocity components is directly set by the wave frequency
$\omega^*$~\cite{Greenspan1968}. For an axisymmetric
distribution of waves, this squared ratio becomes 
$E_z/E_x=2(1-{\omega^*}^2)/(1+{\omega^*}^2)$ \cite{Campagne2015}. In 
Fig.~\ref{fig:comp_anisot_18_nid}, we report the 
componential anisotropy factor $E_z/E_x$ for seven experiments at
$\Omega=18$~rpm and $Re\geq 560$. The ratio $E_z/E_x$ closely follows the
prediction for an axisymmetric distribution of inertial waves over a large 
range of frequencies $0.10<\omega^*<\omega_0^*=0.84$, which corresponds to 
typically $95\%$ of the total kinetic energy of the flow. This 
confirms 
that most of the energy is carried by inertial waves for rapid 
global rotation and high Reynolds number.

Generating such an ensemble of inertial waves is the number one pre-requisite to
achieve weak turbulence in the laboratory. However, it is desirable to also
investigate the validity of the more subtle assumptions of WTT: the large domain
limit -- to avoid purely discrete wave interactions -- and the weak nonlinearity
limit. To wit, we build
on previous studies by Hasselmann {\it et al.}~\cite{Hasselmann} and Aubourg \&
Mordant~\cite{Aubourg2016} and turn to the bicoherence spectrum 
$B(\omega_1,\omega_2)$ of the
horizontal velocity component
\begin{eqnarray}
B=\frac{|\langle\tilde{u}_x(x,z,\omega_1)\tilde{u}_x(x,z,\omega_2)\tilde{u}_x^*(x,z,\omega_1+\omega_2)\rangle_{\bf
 x}|}{\sqrt{ e(\omega_1)e(\omega_2)e(\omega_1+\omega_2)}} \, , \end{eqnarray} 
where 
$^*$ denotes the complex conjugate, $\tilde{u}_x(x,z,\omega)$ is the temporal 
Fourier transform of the horizontal velocity, $\langle\,\rangle_{\bf 
x}$ is a 
spatial average over the measurement field and 
$e(\omega)=\langle|\tilde{u}_x(x,z,\omega)|^2\rangle_{\bf x}$. The bicoherence 
spectrum 
ranges from $B(\omega_1,\omega_2)=0$, when waves at frequencies $\omega_1$, 
$\omega_2$ and $\omega_1+\omega_2$ are uncorrelated, to 
$B(\omega_1,\omega_2)={\cal O}(1)$ when they are perfectly 
phase-correlated~\cite{Aubourg2016}. For instance, in the canonical setup of 
the TRI, a base wave at frequency 
$\omega_0=\omega_1+\omega_2$ transfers energy to waves at frequencies 
$\omega_1$ and $\omega_2$, with a fixed relation between the phases of the 
three waves~\cite{Bordes2012}. We thus expect the bicoherence to be ${\cal 
O}(1)$ for these values of $\omega_1$ and $\omega_2$. Such discrete resonances 
are also the signature of the so-called `discrete wave turbulence' 
regime~\cite{Kartashova2009,Lvov2010,Nazarenko2011,Brouzet2016,LeReun2017,LeReun2019,Davis2020},
where the temporal and/or spatial spectrum remains 
discrete. The framework of WTT departs from such discrete wave turbulence in 
two aspects: first, the large-domain limit, together with the nonlinear 
broadening of the resonances, leads to continuous spectra. Second, wave 
dispersion spontaneously induces a regime where the random-phase approximation 
holds~\cite{Zakharov1992,Newell2011,Nazarenko2011}. More precisely, the 
derivation of the WTT stationary spectrum (\ref{WTTspectrum}) proceeds through 
an expansion in the limit of low Rossby number $Ro$ (based on the injection 
scale and the rms velocity), recalled in the Supplemental Material 
(SM)~\cite{SM}. 
The dominant flow consists of inertial waves, described in terms of helical 
basis vectors~\cite{Cambon89,Smith99} multiplied by slowly varying complex 
amplitudes $b_{s_{\bf i}}^{(0)}$, where ${\bf i}$ denotes the wave vector, the 
polarity $s_{\bf i}=\pm 1$ encodes the sign of the wave helicity, and 
the superscript $(0)$ denotes the lowest-order solution. The latter amplitudes 
have dimension of a velocity, and the phases of the various waves are 
uncorrelated. To lowest order, the numerator of $B$ consists of ensemble 
averages of 
triple products of the form $\langle b_{s_{\bf i}}^{(0)} b_{s_{\bf j}}^{(0)} 
b_{s_{\bf k}}^{(0)\, *} \rangle$, which vanish in the random-phase 
approximation according to Wick's contraction rule~\cite{Nazarenko2011}. One 
needs to consider the next order in the expansion, where smaller contributions 
$b_{s_{\bf i}}^{(1)}$ are forced by quadratic terms in $b_{s_{\bf i}}^{(0)}$. 
The expression of $b_{s_{\bf i}}^{(1)}$ is given in the SM~\cite{SM}, the 
simple order of 
magnitude estimate $b_{s_{\bf k}}^{(1)} \sim k \Omega^{-1} b_{s_{\bf i}}^{(0)\, 
*} b_{s_{\bf j}}^{(0)\, *}$ being sufficient for the present purpose (where the 
right-hand side really is a sum over many such terms for various wavenumbers 
${\bf i}$ and ${\bf j}$ such that ${\bf i}+{\bf j}+{\bf k}={\bf 0}$). A nonzero 
contribution to the numerator of $B$ arises from terms of the form $\langle 
b_{s_{\bf i}}^{(0)} b_{s_{\bf j}}^{(0)} b_{s_{\bf k}}^{(1)\,*} \rangle \sim 
\langle b_{s_{\bf i}}^{(0)} b_{s_{\bf j}}^{(0)} b_{s_{\bf l}}^{(0)} b_{s_{\bf 
m}}^{(0)} \rangle k / \Omega \sim \langle | b_{s_{\bf i}}^{(0)} |^2 \rangle  
\langle | b_{s_{\bf j}}^{(0)} |^2 \rangle k/\Omega$, where we have used Wick's 
contraction rule to transform the quartic term in wave amplitudes into a 
quadratic term in wave intensities. Denoting the generic wave intensity as 
$\langle | b_{s_{\bf i}}^{(0)} |^2 \rangle$, we obtain an estimate for the 
bicoherence $B \sim \langle | b_{s_{\bf i}}^{(0)} |^2 \rangle^2  k/(\Omega  \, 
\langle | b_{s_{\bf i}}^{(0)} |^2 \rangle^{3/2}) \sim \langle | b_{s_{\bf 
i}}^{(0)} |^2 \rangle^{1/2} k/\Omega \sim Ro$. Instead of sharp isolated 
resonance peaks that stand out from a near-zero 
background, the bicoherence $B(\omega_1,\omega_2)$ is now a smooth function 
that takes low ${\cal O}(Ro)$ values.

\begin{figure}
	\centering{\includegraphics[width=8.5cm]{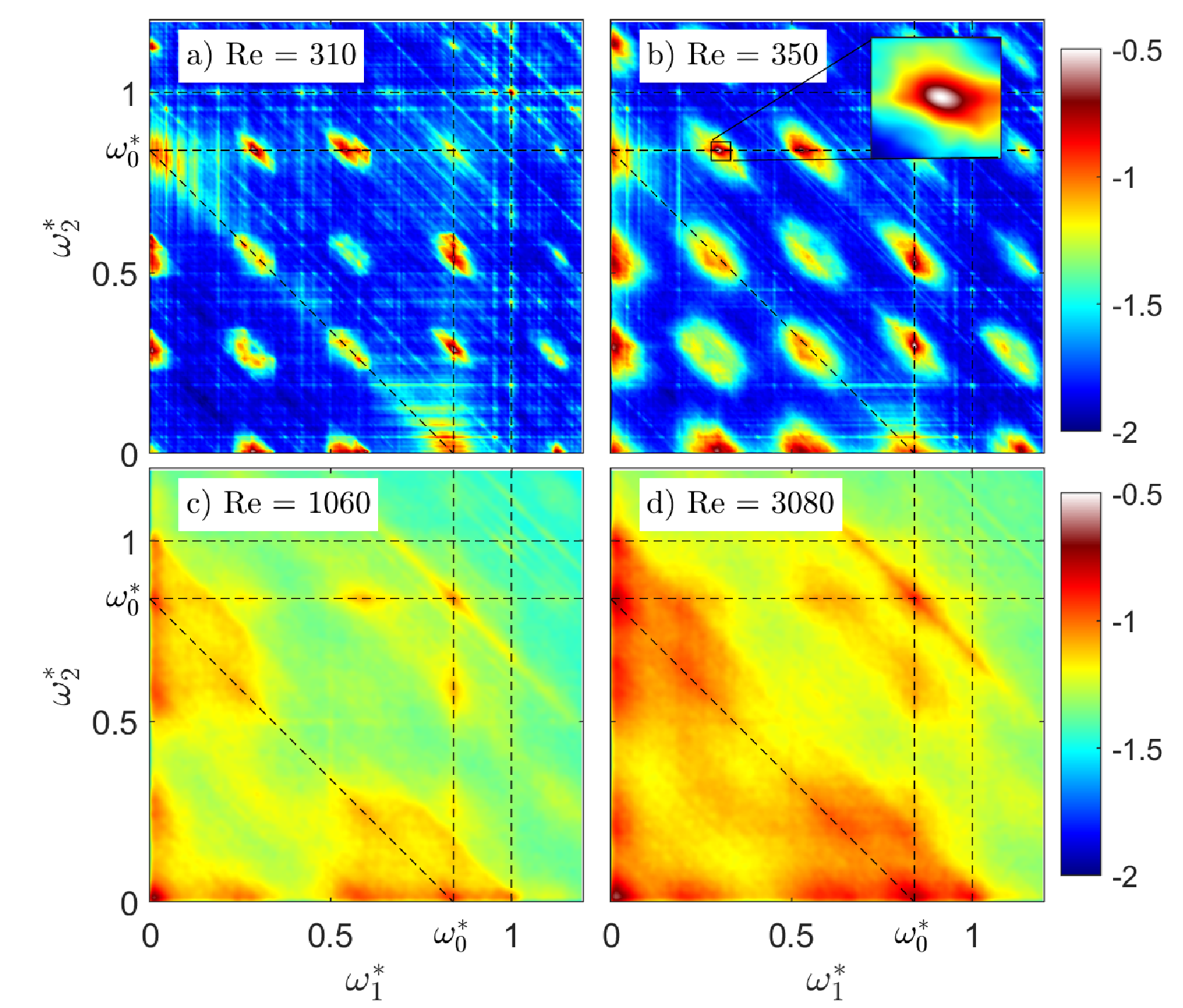}}
	\caption{Logarithm of the bicoherence $B$ for $\Omega=18$~rpm and 
		increasing values of $Re$. \label{fig:bicoherence_18_nid}}
\end{figure}

In Fig.~\ref{fig:bicoherence_18_nid}, we show the experimental bicoherence $B$ 
for four values of $Re$ at $\Omega=18$~rpm. For
$Re=310$, slightly above the threshold of the TRI, the
bicoherence map consists in an array of peaks localized at all coordinate 
values 
$(\omega_1^*, \omega_2^*)$ associated with two of the three energetic
frequencies: $\omega^*\simeq 0.29$, $\omega^* \simeq 
0.55$ and $\omega_0^*=0.84$. This is the signature of the TRI of the base 
waves, which induces a regime of discrete 
wave interactions, as described above. 
Further from the threshold of 
the first 
triadic instability, for $Re=350$, one notices the nonlinear broadening of the 
resonance peaks in the bicoherence map. At large distance from the TRI 
threshold, 
for 
$Re=3080$, the bicoherence has become a smooth function that takes low values 
ranging from $5\times 10^{-2}$ to $10^{-1}$, comparable to the Rossby number
based on the 
rms velocity ($1.7$~cm/s) and the injection wavelength ($14$~cm inside the PIV 
plane), $Ro\simeq 3 \times 
10^{-2}$. The experimental bicoherence thus confirms the gradual transition from
a discrete-wave-interaction regime to a proper weak turbulence regime 
as $Re$ increases. In the latter regime, the 
discreteness of the 
modes is smoothed out by the nonlinear broadening of the resonances, and both 
the temporal spectrum and the bicoherence become smooth functions. The 
bicoherence settles at a low value, of order 
$Ro$, compatible with a weakly 
nonlinear wave field that satisfies the random phase approximation.

\begin{figure}
	\centering{\includegraphics[width=8cm]{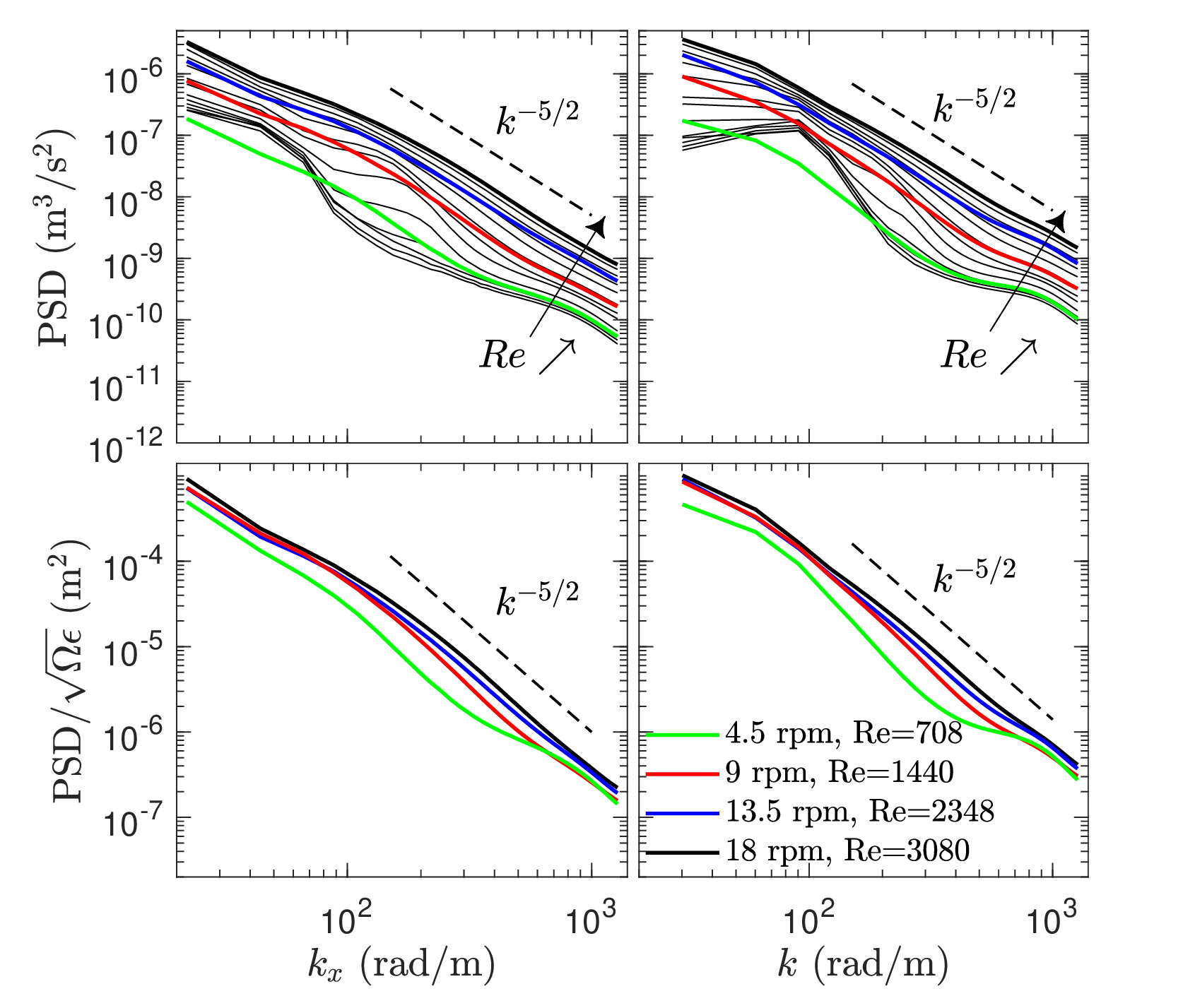}}
	\caption{Spatial energy spectrum as a function of the horizontal 
	wavenumber $k_x$ (left) and wavenumber $k=\sqrt{k_x^2+k_z^2}$ (right). 
	\textbf{Top 
	panels:} black thin lines correspond to $\Omega=18$~rpm and	$Re$ increasing 
	from $270$ to $3080$. Thick lines correspond to the highest Reynolds 
	number for 
	various rotation rates (see legend). \textbf{Bottom panels:} when rescaled 
	by the spectral level predicted by WTT, the highest-$Re$ spectra at large 
	$\Omega$ collapse onto a master curve. Dashed lines show the WTT power-law 
	prediction. \label{fig:spatial_spectrum_18_nid}}
\end{figure}

\textit{Cascading states.---} Having established that the turbulent flow is
compatible with the assumptions of WTT at large $Re$, we turn to the spatial
energy spectrum, with the goal of testing the predictions of WTT. We first
compute the 2D spatial spectrum of the PIV velocity fields. In the left panels of
Fig.~\ref{fig:spatial_spectrum_18_nid}, we integrate this 2D spectrum over the
vertical wavenumber $k_z$ and show the resulting 1D spectrum as a function of
$k_x$. In the right panels, we integrate the 2D spectrum over the angular
direction instead, before plotting the resulting 1D spectrum as a function of
the wavenumber $k=\sqrt{k_x^2+k_z^2}$. All the spectra in
Fig.~\ref{fig:spatial_spectrum_18_nid} are normalized in such a way that the
integral of the spectrum over its variable ($\int {\rm PSD}(k_x)\,dk_x$ or 
$\int 
{\rm PSD}(k)\,dk$) yields the space- and
time-averaged kinetic energy inside the PIV domain. In the top panels, for weak 
driving amplitude, the spectra
display a bump at a wavenumber corresponding to the injection wavelength. As 
the driving amplitude
increases, the nonlinearities populate higher and higher wavenumbers in the
spectrum, up to the point where a self-similar cascade develops: the high-$Re$
spectra then display a power-law behavior, with an exponent in close agreement 
with the prediction $-5/2$ of Eq.~(\ref{WTTspectrum}). This is
clearly visible when the spectrum is shown as a function of $k_x$, for which the
prediction is made, but also when the spectral content is shown as a function of
$k$.

Beyond the prediction of the spectral exponent, WTT provides a prediction for 
the spectral level as a function of the rotation rate $\Omega$, the mean energy 
dissipation rate $\epsilon$, and the vertical wavelength $L_\parallel$ of the 
forced waves. In the highest-$Re$ experiments for the four values of $\Omega$ 
in Fig.~\ref{fig:spatial_spectrum_18_nid},
$L_\parallel$ is nearly constant and equal to $16$~cm. A direct 
measurement of $\epsilon$ is 
a notoriously difficult task, 
that requires well-resolved fully 3D velocity fields. The present PIV data are 
well-resolved but 
2D. Assuming statistical axisymmetry, a proxy for $\epsilon$ can be obtained 
from the velocity gradients accessible in the measurement plane:
\begin{equation*}
\epsilon \simeq \nu \langle
2 \left(\partial_x u_x\right)^2
+ 2 \left(\partial_x u_z\right)^2 + 2 \left(\partial_z u_x\right)^2
+ \left(\partial_z u_z\right)^2
\rangle_{{\bf x},t},
\end{equation*}
where $\langle \, \rangle_{{\bf x},t}$ is a spatial and temporal average.
Using this proxy, we plot in 
Fig.~\ref{fig:spatial_spectrum_18_nid} the rescaled spectra 
${\rm PSD}/\sqrt{\Omega 
\epsilon}$, for the highest Reynolds number achieved at each rotation 
rate. This representation leads to a collapse 
of the high-$\Omega$ spectra 
onto a master curve that follows the $k^{-5/2}$ power-law dependence. The 
experimental data thus validate the predictions of WTT, both for the spectral 
level and the spectral exponent, provided molecular dissipation is negligible 
(high $Re$) and the wave turbulence is weakly nonlinear (high $\Omega$, low 
$Ro$).

\textit{Discussion.---} The present experimental apparatus allows us to 
generate a turbulent flow that consists of weakly interacting inertial waves in 
a large fluid domain. As the forcing increases, the system transitions from a 
regime of discrete wave interactions to a regime that displays continuous 
temporal spectra and bicoherence maps, in line with WTT. At high Reynolds 
number and low Rossby number, the resulting spatial spectrum exhibits the 
scaling properties predicted by WTT, both in terms of spectral slope and 
spectral level. Such a laboratory realization of weak turbulence in a 3D fluid 
system could open an experimental avenue for studies that gradually incorporate 
the additional complexities of natural flows. Among the many exciting 
directions for future research, one could add density stratification to 
characterize the turbulent mixing induced by inertia-gravity waves in the weak-turbulence 
regime~\cite{Lvov2001}, and one could progressively relax the damping of the 
geostrophic flow to characterize its impact on the wave-turbulent 
dynamics. Following Scott~\cite{Scott2014}, it may be that the large-scale 
geostrophic flow sweeps the wave phases and challenges a precise 
characterization of the wave dynamics, but that the small-scale cascading 
dynamics remains largely unaffected. The consequence would be that WTT remains 
a valuable tool to charaterize small-scale dissipation and to develop 
subgrid-scale parameterizations in that context.

\begin{acknowledgments}
We acknowledge J. Amarni, A. Aubertin, L. Auffray
and R. Pidoux for experimental help. This work was supported by a grant from 
the Simons Foundation (651461, PPC) and by the Agence Nationale de la Recherche 
through Grant ``DisET'' No.~ANR-17-CE30-0003. BG acknowledges support by
the European Research Council under grant agreement 757239.
\end{acknowledgments}

\end{document}